\documentstyle[prl,aps,preprint]{revtex}
\makeatletter
\def\section{\@mainheadtrue
\@startsection {section}{1}{\z@}{0.8cm plus1ex minus .2ex}
               {0pt}{\reset@font\bf}}
\def\@sect#1#2#3#4#5#6[#7]#8{\ifnum #2>\c@secnumdepth
\let\@svsec\@empty\else
\refstepcounter{#1}%
\def\@tempa{#8}%
\ifx\@tempa\empty %
\ifappendixon %
\if@mainhead %
\def\@tempa{APPENDIX }\def\@tempb{}%
\else %
\def\@tempa{}\def\@tempb{. \enskip\enskip }%
\fi
\else %
\def\@tempa{}\def\@tempb{. \enskip\enskip }%
\fi
\else %
\ifappendixon %
\if@mainhead %
\def\@tempa{APPENDIX }\def\@tempb{: }%
\else %
\def\@tempa{}\def\@tempb{. \enskip\enskip }%
\fi
\else %
\def\@tempa{}\def\@tempb{. \enskip\enskip }%
\fi
\fi
\edef\@svsec{\@tempa\csname the#1\endcsname\@tempb}\fi
\@tempskipa #5\relax
\ifdim \@tempskipa>\z@
\begingroup #6\relax
{\hskip #3\relax\@svsec}{\interlinepenalty \@M
\if@mainhead\uppercase{#8}\else#8\fi\par}%
\endgroup
\csname #1mark\endcsname{#7}\addcontentsline
{toc}{#1}{\ifnum #2>\c@secnumdepth \else
\protect\numberline{\csname the#1\endcsname}\fi
#7}\else
\def\@svsechd{#6\hskip #3\relax %
\@svsec \if@mainhead\uppercase{#8}\else#8\fi
\csname #1mark\endcsname
{#7}\addcontentsline
{toc}{#1}{\ifnum #2>\c@secnumdepth \else
\protect\numberline{\csname the#1\endcsname}\fi
#7}}\fi
\@xsect{#5}}
\makeatother

\begin{document}

\title{Another View on Massless Matter-Gravity Fields \\
in Two Dimensions\thanks
{This work is supported in part
by funds provided by the U.S.
Department of Energy (D.O.E.) under cooperative
agreement \#DF-FC02-94ER40818}}

\author{R. Jackiw}

\address{{~}\\ \baselineskip=14pt
Center for Theoretical Physics,
Laboratory for Nuclear Science
and Department of Physics \\
Massachusetts Institute of Technology \\
Cambridge, Massachusetts~ 02139 \\ {~} }

\date {MIT-CTP: \#2377 \hfill  HEP-TH: /9501016
                       \hfill  January 1995}

\maketitle

\begin{abstract}
Conventional quantization of two-dimensional diffeomorphism and Weyl
invariant theories sacrifices the latter symmetry to anomalies, while
maintaining the former.  When alternatively Weyl invariance is
preserved by abandoning diffeomorphism invariance, we find that some
invariance against coordinate redefinition remains: one can still
perform at will transformations possessing a constant Jacobian.  The
alternate approach enjoys as much ``gauge'' symmetry as the
conventional formulation.
\end{abstract}

\bigskip\bigskip

\section{}
The theory of a massless scalar field $\phi$, interacting with
2-dimensional gravity that is governed solely by a metric tensor
$g_{\mu\nu}$ has a conventional description: functionally integrating
$\phi$ produces an effective action $\Gamma^P$, a functional of
$g^{\mu\nu}$, which has been given by Polyakov as~\cite{ref:1}
\begin{equation}
\Gamma^P (g^{\mu\nu}) = \frac{1}{96\pi}
           \int d^2 x  \, d^2 y \sqrt{-g(x)} \,
           R(x) K^{-1}(x,y) \sqrt{-g(y)} R(y)
\label{eq:1}
\end{equation}
Here $R$ is the scalar curvature and $K^{-1}$ satisfies
\begin{equation}
 - \frac{1}{\sqrt{-g(x)}}  \frac{\partial}{\partial x^\mu} \sqrt{-g(x)} \,
           g^{\mu\nu}(x)  \frac{\partial}{\partial x^\nu} K^{-1}(x,y) =
           \frac{1}{\sqrt{-g(x)}} \delta^2 (x-y)
\label{eq:2:}
\end{equation}
Eq.~(\ref{eq:1}) results after definite choices are made to resolve
ambiguities of local quantum field theory: it is required that
$\Gamma^P$ be diffeomorphism invariant and lead to the conventional
trace (Weyl) anomaly.  This translates into the conditions that the
energy-momentum tensor
\begin{equation}
 \Theta^{P}_{\mu\nu} = \frac{2}{\sqrt{-g}}
           \frac{\delta\Gamma^P} {\delta g^{\mu\nu}}
\label{eq:3}
\end{equation}
be covariantly conserved (diffeomorphism invariance of $\Gamma^P$),
\begin{equation}
  D_\mu \left(g^{\mu\nu}\Theta_{\nu\alpha}^P\right) = 0
\label{eq:4}
\end{equation}
and possess a non-vanishing trace (Weyl anomaly).
\begin{equation}
  g^{\mu\nu} \Theta^{P}_{\mu\nu} = \frac{1}{24\pi} R
\label{eq:5}
\end{equation}
Equations (\ref{eq:3})--(\ref{eq:5}) can be integrated to give (\ref{eq:1});
also from (\ref{eq:1}) and (\ref{eq:3})
one finds that\footnote{%
When the gravity field $g_{\mu\nu}$ is viewed
as externally prescribed, $\Theta^P_{\mu\nu}$ is the vacuum matrix element
of the operator energy-momentum tensor for the quantum field $\phi$.
Eq.~(\ref{eq:6})
has been derived by M. Bos [2], not by varying the Polyakov action
\cite{ref:1}, but by direct computation of the relevant expectation value.
}  %% end footnote
\begin{eqnarray}
  \Theta^{P}_{\mu\nu} & = & -\frac{1}{48\pi}
           \left( \partial_\mu \Phi \partial_\nu \Phi
                - \frac{1}{2} g_{\mu\nu} g^{\alpha\beta} \partial_\alpha \Phi
                      \partial_\beta\Phi \right)
           \nonumber \\
  &  & \quad - \frac{1}{24\pi}
           \left( D_\mu D_\nu \Phi - \frac{1}{2}  g_{\mu\nu}
                      g^{\alpha\beta} D_\alpha D_\beta \Phi \right)
           +  \frac{1}{48\pi}  g_{\mu\nu} R
\label{eq:6}
\end{eqnarray}
where $\Phi$ is the solution to
\begin{equation}
  g^{\alpha\beta} D_\alpha D_\beta \Phi = \frac{1}{\sqrt{-g}} \,
           \partial_\alpha \sqrt{-g} \, g^{\alpha\beta} \partial_\beta \Phi = R
\label{eq:6b}
\end{equation}
One easily verifies that (\ref{eq:6}) obeys (\ref{eq:4}) and (\ref{eq:5}).
Notice that the traceless
part of $\Theta^P_{\mu\nu}$ satisfies
\begin{equation}
  D_\mu \Bigl( g^{\mu\nu} \Theta^P_{\nu\alpha}
           \Bigm|_{\raisebox{-.2ex}{$\scriptstyle\rm traceless$}} \, \Bigr)
           = \frac{1}{48\pi} \partial_\alpha {\it R}
\label{eq:7}
\end{equation}

It is well known that one can make alternative choices when defining
relevant quantities.  In particular, one can abandon diffeomorphism
invariance and obtain an alternate effective action $\Gamma$, which is
Weyl invariant because it is a functional solely of the Weyl invariant
combination\footnote{%
See Ref.~\cite{ref:5}.  A point of view that provides another
alternative to Polyakov's approach has recently
appeared in Ref.~\cite{ref:2}.
} %%end of footnote
\begin{equation}
        \gamma^{\mu\nu} \equiv \sqrt{-g} \, g^{\mu\nu}
\label{eq:8}
\end{equation}
This ensures vanishing trace for the modified energy momentum tensor.
\begin{eqnarray}
  \Theta_{\mu\nu}
           & = & \frac{2}{\sqrt{-g}}
                      \frac{\delta\Gamma}{\delta g^{\mu\nu}}
                      = 2 \frac{\delta\Gamma}{\delta\gamma^{\mu\nu}}
                      -  \gamma_{\mu\nu} \gamma^{\alpha\beta}
                       \frac{\delta\Gamma}{\delta\gamma^{\alpha\beta}}
           \label{eq:9} \\
  \gamma^{\mu\nu} \Theta_{\mu\nu}
           & = & 0
           \label{eq:10}
\end{eqnarray}
Here $\gamma_{\mu\nu}$ is the matrix inverse to $\gamma^{\mu\nu}$,
\begin{equation}
  \gamma_{\mu\nu} = g_{\mu\nu} / \sqrt{-g}
\label{eq:11}
\end{equation}
and $ {\det} \gamma_{\mu\nu} = {\det} \gamma^{\mu\nu} = -1$.

In this Letter we study more closely the response of $\Gamma\/$ to
diffeomorphism transformations when Weyl symmetry is preserved.  We
find that diffeomorphism invariance is not lost completely; rather it
is reduced: $\Gamma\/$ remains invariant against transformations that
possess a constant (unit) Jacobian --- we call this $S$-diffeomorphism
invariance.\footnote{%
$S$-diffeomorphisms preserve local area
$\sqrt{-g} \, d^2x$ on spaces where $\sqrt{-g}$ is constant.  (I thank
W.~Taylor and B.~Zwiebach for discussions on this.)
} %% end of footnote
In the absence of diffeomorphism invariance,
$\Theta_{\mu\nu}$ is no longer covariantly conserved; nevertheless we
shall show that $S$-diffeomorphism invariance restricts the divergence
of $\Theta_{\mu\nu}$ [essentially to the form given in (\ref{eq:7})].
We shall argue that our alternative evaluation follows the intrinsic
structures of the theory more closely than the conventional approach.

\section{}
Before presenting our argument, we define notation and record some formulas.
The 2-dimensional Euler density is a total derivative.
\begin{equation}
  \sqrt{-g} R = \partial_\mu R^\mu \>, \qquad
           R = D_\mu \left( R^\mu / \sqrt{-g} \right)
\label{eq:12}
\end{equation}
But $R^\mu$  cannot be presented explicitly and locally in terms of the metric
tensor and its derivatives
as a whole; rather it is necessary to parametize $g_{\mu\nu} =
\sqrt{-g}\gamma_{\mu\nu}$.  We define
\begin{mathletters}%%
\begin{equation}
    \sqrt{-g} = e^\sigma
\label{eq:13a}
\end{equation}
and parametize the light-cone $[(\pm) \equiv \frac{1}{\sqrt{2}} (0 \pm 1)]$
components of $\gamma_{\mu\nu}$ as
\begin{eqnarray}
  \gamma_{++} & = & -\gamma^{--} = e^\alpha {\sinh} \beta \nonumber \\
  \gamma_{--} & = & -\gamma^{++} = e^{-\alpha} {\sinh} \beta \nonumber \\
  \gamma_{+-} & = & \gamma_{- +} = \gamma^{+ -}
           = \gamma^{- +} =  {\cosh} \beta
\label{eq:13b}%%
\end{eqnarray}
\label{eq:13ab}%%
\end{mathletters}%%
Then the formula for $R^\mu$ reads
\begin{equation}
  R^\mu = \gamma^{\mu\nu} \partial_\nu \sigma +
  \partial_\nu\gamma^{\mu\nu}-\epsilon^{\mu\nu} ({\cosh} \beta-1)
  \partial_\nu\alpha
\label{eq:14}
\end{equation}
where the explicit parametrization (\ref{eq:13ab}) is needed to
present the last term in (\ref{eq:14}).\footnote{%
This is analogous to what happens with a Chern-Simons term.
Upon performing a gauge
transformation with a gauge function $U$, the Chern-Simons term
changes by a total derivative.  However, direct evaluation of the
gauge response includes the expression $\omega = \frac{1}{24\pi^2}
{\rm tr} \epsilon^{\alpha\beta\gamma} U^{-1} \partial_\alpha U
U^{-1}\partial_\beta U^{-1}\partial_\gamma U$, which can be recognized
as a total derivative only after $U$ is explicitly parametrized.  For
example, in SU(2) $U = { \exp } \,\theta, \, \theta = \theta^a\sigma^a
/2i$, and $\omega= \partial_\alpha
\omega^\alpha$
where $\omega^\alpha = \frac{1}{4\pi^2} {\rm tr} \epsilon^{\alpha\beta\gamma}
\theta\partial_\beta \theta\partial_\gamma \theta
\left(\frac{\mid\theta\mid - {\sin} \mid\theta\mid}
{|\theta|^3}\right)$ with
$|\theta| \equiv \sqrt{\theta^a\theta^a}$; see Jackiw in~\cite{ref:4}.
} %%end of footnote
(Here $\epsilon^{\mu\nu}$ is the anti-symmetric
numerical quantity, normalized by $\epsilon^{01} = 1$.)

Even though the last contribution in (\ref{eq:14}) to $R^\mu\/$ is not
expressible in terms of $g_{\mu \nu}\/$ or $\gamma_{\mu \nu}\/$, its
arbitrary variation satisfies a formula involving only $\gamma_{\mu \nu}\/$.
\begin{equation}
\delta \Bigl[ \epsilon^{\mu \nu} (\cosh \beta - 1) \partial_\nu \alpha \Bigr]
     - \partial_\nu \Bigl[ \epsilon^{\mu \nu} (\cosh \beta - 1)
           \delta \alpha \Bigr]
     = \frac{-1}{2} \gamma^{\mu \nu}
        \Bigl( \partial_\alpha \gamma_{\nu \beta}
           + \partial_\beta \gamma_{\nu \alpha}
        - \partial_{\nu} \gamma_{\alpha \beta} \Bigr)
        \delta \gamma^{\alpha \beta}
\label{eq:14a}
\end{equation}
Note that the right side equals
$- \gamma^\mu_{\alpha \beta} \delta \gamma^{\alpha \beta}\/$,
where $\gamma^\mu_{\alpha \beta}\/$
is the Christoffel connection when the metric tensor
is $\gamma_{\mu \nu} : \gamma^{\mu}_{\alpha \beta}
= \Gamma^{\mu}_{\alpha \beta}
\biggm|_{\stackrel{\cdot}{g_{\mu \nu} = \gamma_{\mu \nu}}}\/$

While the covariant divergence of $R^\mu / \sqrt{-g}\/$ is the scalar
curvature, see (\ref{eq:12}), $R^\mu / \sqrt{-g}\/$ does not transform
as a vector under coordinate redefinition.  Rather for an
infinitesimal diffeomorphism generated by $f^\mu\/$
\begin{equation}
\delta_D x^\mu = - f^\mu (x)
\label{eq:14b}
\end{equation}
one verifies that
\begin{mathletters}%%
\begin{equation}
\delta_D (R^\mu / \sqrt{-g})
           = L_f  (R^\mu / \sqrt{-g})
           + \frac{1}{\sqrt{-g}} \epsilon^{\mu \nu} \partial_\nu \Delta_{f}
\label{eq:14c1}
\end{equation}
where $L_f\/$ in the Lie derivative with respect to $f^\mu\/$, and
\begin{equation}
\Delta_f \equiv
           ( \partial_{+} - e^\alpha \tanh \frac{\beta}{2} \partial_{-} ) f^{+}
           - ( \partial_{-} - e^{- \alpha} \tanh \frac{\beta}{2}
                      \partial_{+} ) f^{-}
\label{eq:14c2}
\end{equation}
\label{eq:14c}%%
\end{mathletters}%%
This non-tensorial transformation rule nevertheless ensures a scalar
transformation law for $D_\mu (R^\mu / \sqrt{-g})\/$.  Consequently, a
world scalar action may be constructed by coupling vectorially
$R^\mu\/$ to a scalar field $\Psi\/$, $I_V = \int d^2 x \, R^\mu
\partial_\mu \Psi\/$; invariance is verified from (\ref{eq:12}) after
partial integration.  An axial coupling also produces a world scalar
action, $I_A = \int \frac{d^2 x}{\sqrt{-g}} \, R_\mu \epsilon^{\mu
\nu} \partial_\nu \Psi\/$, provided $\Psi\/$ satisfies $g^{\mu \nu}
D_\mu D_\nu \Psi = 0\/$; this follows from~(\ref{eq:14c}).

Finally we remark that the last term in (\ref{eq:14}) naturally
defines a 1-form $a \equiv (\cosh \beta - 1) d \alpha\/$ and the
2-form $\omega = d a = \sinh \beta d \beta d \alpha\/$.  These are
recognized as the canonical 1- form and the symplectic 2-form,
respectively, for SL $(2, R)$.  Indeed $\omega\/$ also equals
$\frac{1}{2} \epsilon_{a b c} \xi^a d \xi^b d \xi^c\/$, where
$\xi^a\/$ is a three vector on a hyperboloid = SL $(2, R) / U (1) :
(\xi^1)^2 - (\xi^2)^2 - (\xi^3)^2 = -1$.  Effectively, $\omega\/$ is
the Kirillov-Kostant 2-form on SL $(2,R)$.\footnote{%
I thank V. P. Nair for pointing this out.  }  %% end of footnote

\section{}
The Lagrange density for our theory reads
\begin{equation}
   {\cal L} = \frac{1}{2} \sqrt{-g} g^{\mu\nu} \partial_\mu \phi
  \partial_\nu \phi =
  \frac{1}{2} \gamma^{\mu\nu} \partial_\mu \phi \partial_\nu \phi
\label{eq:15}
\end{equation}
Equivalently, a first-order expression may be given,
\begin{equation}
 {\tilde{\cal L}} = \Pi \dot{\phi} - u {\cal E} - v{ P}
\label{eq:16}
\end{equation}
where ${\cal E}$ and ${\cal P}$ are the free-field energy and momentum
densities
\begin{mathletters}%%
\begin{eqnarray}
  {\cal E} & = & \frac{1}{2} \Pi^2 + \frac{1}{2} (\phi')^2
                      \label{eq:17a} \\
{\cal P} & = & - \phi'\Pi
                      \label{eq:17b}
\end{eqnarray}
\label{eq:17ab}%%
\end{mathletters}%%
Here dot and dash signify time ($x^0 \equiv t$) and space ($x^1 \equiv
x$) differentiation, respectively.  The gravitational variables enter
as Lagrange multipliers, in $\tilde{\cal L}$
\begin{eqnarray}
  u & = & \frac{1}{\sqrt{-g} g^{00}} =  \frac{1}{\gamma^{00}}
                    \nonumber \\
  v & = & \frac{g^{01}}{g^{00}} = \frac{\gamma^{01}}{\gamma^{00}}
\label{eq:18}
\end{eqnarray}
enforce vanishing ${\cal E}$ and ${\cal P}$.  It is seen that only two
of the three independent components in $g_{\mu\nu}$ are present:
$\sigma={\ln} \sqrt{-g}$ does not occur in ${\cal L}$ or $\tilde{\cal L}$,
which depend only on $\gamma^{\mu\nu}$ --- this is of course a
manifestation of Weyl invariance.

In spite of the absence of $\sigma$ in the classical theory,
Polyakov's quantum effective action (\ref{eq:1}) carries a
$\sigma$-dependence.  The breaking of Weyl symmetry arises when one
evaluates the functional determinant that leads to the effective
action; {\it viz.}~$-\frac{1}{2} {\ln} \det K$, where $K$ is the
kernel present in the classical action.
\begin{equation}
  K(x,y) = - \frac{\partial}{\partial x^\mu} \gamma^{\mu\nu}
  (x) \frac{\partial}{\partial x^\nu} \delta^2(x-y)
\label{eq:19}
\end{equation}
Formally the determinant is given by the product of K's eigenvalues,
$\det K = \Pi_\lambda \lambda$, but it still remains to formulate the
eigenvalue problem.  The diffeomorphism invariant definition
recognizes that $K$ is a density, so eigenvalues are defined by
\begin{mathletters}%%
\begin{eqnarray}
  \int K \Psi^P_\lambda
           &=&  \sqrt{-g} \lambda \Psi_\lambda^P
           \nonumber \\
  - g^{\alpha\beta} D_\alpha D_\beta \Psi_\lambda^P
           &=& \lambda \Psi_\lambda^P
\label{eq:20a}
\end{eqnarray}
and the inner product involves an invariant measure
\begin{equation}
  \left<\lambda_1\mid\lambda_2\right>^P = \int \sqrt{-g}
  \Psi_{\lambda_1}^{P*} \Psi^P_{\lambda_2}
\label{eq:20b}
\end{equation}
\label{eq:20ab}%%
\end{mathletters}%%
In this way $\sigma = {\ln}\sqrt{-g}$ enters the calculation.

However, one may say that it is peculiar to introduce into the
determination of eigenvalues a variable that is not otherwise present
in the problem.  (Below we shall also argue that it is unnatural to
insist on diffeomorphism invariance.)

As an alternative to (\ref{eq:20ab}) one may define eigenvalues without
inserting~$\sigma$,
\begin{mathletters}%%
\begin{equation}
  \int K\Psi_\lambda = \lambda\Psi_\lambda
\label{eq:21a}
\end{equation}
and use a $\sigma$-independent inner product.
\begin{equation}
  \left<\lambda_1\mid\lambda_2\right> = \int \Psi^*_{\lambda_1}
  \Psi_{\lambda_2}
\label{eq:21b}
\end{equation}
\label{eq:21ab}%%
\end{mathletters}%%
It follows that the effective action will be as in (\ref{eq:1}), with
$\sigma$ set to zero.
\begin{equation}
  \Gamma(\gamma^{\mu\nu})=\frac{1}{96\pi}\int d^2 x d^2 y
  {\cal R}(x) K^{-1}(x,y)
  {\cal R}(y)
\label{eq:22}
\end{equation}
Here ${\cal R}$ is the scalar curvature computed with $\gamma^{\mu\nu}
   (\gamma_{\mu\nu})$ as the contravariant (covariant) metric tensor.
   From (\ref{eq:12}) -- (\ref{eq:14}) we have
\begin{eqnarray}
  {\cal R} & = & \partial_\mu {\cal R}^\mu
                      \label{eq:23} \\
  {\cal R}^\mu & = & \partial_\nu \gamma^{\mu\nu} - \epsilon^{\mu\nu}
           ({\cosh}\beta-1) \partial_\nu \alpha
\label{eq:24}
\end{eqnarray}
Evidently $\Gamma$ is a functional solely of $\gamma^{\mu\nu}$; since
it does not depend on $\sigma$ it is Weyl invariant, leading to a
traceless energy-momentum tensor as in~(\ref{eq:10}).

Of course the definitions (\ref{eq:21ab}) do not respect
diffeomorphism invariance; however they are invariant against
S-diffeomorphisms.  Consequently $\Gamma$ also is $S$-diffeomorphism
invariant.  With the help of (\ref{eq:14}), (\ref{eq:22}) and
(\ref{eq:24}) we can exhibit the relation between $\Gamma^P$ and
$\Gamma$.  Using (\ref{eq:12}) and (\ref{eq:14}) to evaluate
(\ref{eq:1}), and integrating by parts the terms involving $\sigma$ to
remove the non-local kernel $K^{-1}$, leaves
\begin{equation}
  \Gamma^P (g^{\mu\nu}) = \frac{1}{96\pi} \int d^2 x \partial_\mu  \sigma
  \gamma^{\mu\nu} \partial_\nu \sigma + \frac{1}{48\pi} \int d^2 x
  \partial_\mu \sigma {\cal R}^\mu + \Gamma (\gamma^{\mu\nu})
\label{eq:25}
\end{equation}
Thus the diffeomorphism invariance restoring terms, present in
$\Gamma^P$, add to $\Gamma$ a local expression, which is a quadratic
polynomial in~$\sigma$.  The locality of $\Gamma^P - \Gamma$
highlights its arbitrariness, but $\Gamma$ has the advantage of not
involving quantities extraneous to the problem.
[Formula~(\ref{eq:25}) may also be presented as $\Gamma^P (g^{\mu \nu}) =
\frac{1}{96 \pi} \int (\sigma - K^{-1} {\cal R}) K  (\sigma - K^{-1}
{\cal R})\/$.]

Infinitesimal coordinate transformations make use of two arbitrary
functions $f^\mu\/$, see (\ref{eq:14b}).  $S$-diffeomorphisms possess
unit Jacobian, so infinitesimally $\partial_\mu f^\mu = 0$;
consequently only one function survives.
\begin{equation}
  \delta_{SD} x^\mu = \epsilon^{\mu\nu} \partial_\nu f(x)
\label{eq:27}
\end{equation}
Since Weyl transformations
\begin{equation}
  g_{\mu\nu} \rightarrow e^W g_{\mu\nu}
\label{eq:28}
\end{equation}
also make use of a single function, replacing diffeomorphism
invariance, involving two arbitrary functions $f^\mu$, by Weyl and
$S$-diffeomorphism invariance still leaves two arbitrary functions,
$f$ and $W$.  Indeed, similar to diffeomorphism invariance, the
combination of Weyl and $S$-diffeomorphism invariance can be used to
reduce a generic metric tensor, containing three functions, to a
single arbitrary function.

In particular by using their respective symmetries, we can bring
$\Gamma^P(g^{\mu\nu})$ and $\Gamma (\gamma^{\mu\nu})$ into equality.
Diffeomorphism invariance allows placing $g_{\mu\nu}$ into the
light-cone gauge, where $g_{--} = 0, \, g_{+-} = 1$ and $g_{++}$ is
the arbitrary function $h_{++}$ \cite{ref:1}.  Correspondingly, with
$S$-diffeomorphism invariance we can set to zero the $(--)$ component
in $\gamma_{\mu\nu}$ and the $(+-)$ component to unity.  This is
achieved by passing from the original variables $\left\{x^\mu \right\}$
and metric function $\gamma_{\mu\nu}(x)$ to a new quantities
$\left\{\tilde{x}^\mu \right\}$ and $\tilde{\gamma}_{\mu v}
(\tilde{x})$, where
\begin{mathletters}%%
\begin{eqnarray}
  \frac{\partial x^+}{\partial \tilde{x}^-}
           & = & - \frac{\gamma_{--}}{\gamma_{+-}\pm 1}
                                            \label{eq:29a}  \\
\frac{\partial x^+}{\partial \tilde{x}^+}
           & = &  \frac{-\gamma_{--}}{\gamma_{+-}\pm 1}
           \frac{\partial x^-}{\partial \tilde{x}^-} +
           \frac{c}{\partial x^{-} / \partial \tilde{x}^- }
                      \nonumber
\end{eqnarray}
Either sign may be taken in $\gamma_{+-} \pm 1$ and $c^2=1$.
One then finds
\begin{eqnarray}
 \tilde{\gamma}_{--}
           & = & 0,
                      \qquad\qquad
                   \tilde{\gamma}_{+-} = 1
                       \nonumber \\
\tilde{\gamma}_{++} (\tilde{x})
           & = & \frac{\gamma_{++}(x)}
                                 {(\partial x^-/\partial \tilde{x}^-)^2} +
           2c \, \frac{\partial x^{-} / \partial \tilde{x}^{+}}
                                 {\partial x^{-} / \partial \tilde{x}^{-}}
                                 \label{eq:29b}
\end{eqnarray}
\label{eq:29ab}%%
\end{mathletters}%%
Upon identification of $\tilde{\gamma}_{++}$ with $h_{++}$,
$\Gamma^P = \Gamma$ in the selected gauge.

Under an infinitessimal diffeomorphism
\begin{equation}
  \delta_D\Gamma = \int d^2 x \sqrt{-g} f^\alpha D_\mu
                               \Theta^\mu_{\;\; \alpha}
\label{eq:31}
\end{equation}
so it follows from (\ref{eq:27}) that for $S$-diffeomorphisms
\begin{equation}
  \delta_{SD}\Gamma = \int d^2 x f \epsilon^{\alpha\beta}
  \partial_\beta \left(\sqrt{-g} D_\mu\Theta^\mu_{\;\;\alpha}\right)
\label{eq:32}
\end{equation}
and invariance is equivalent to vanishing of the integrand.  But
$\sqrt{-g} D_\mu \Theta^\mu_{\;\;\alpha} = \break
 \partial_\mu(\sqrt{-g}g^{\mu\nu}\Theta_{\nu\alpha}) + \frac{\sqrt{-g}}{2}
  \partial_\alpha g^{\mu\nu}\Theta_{\mu\nu}$, which for traceless
$\Theta_{\mu\nu}$ may be written as
$\partial_\mu (\gamma^{\mu\nu}\Theta_{\nu\alpha}) + \frac{1}{2} \partial_\alpha
  \gamma^{\mu\nu} \Theta_{\mu\nu} = d_\mu (\gamma^{\mu\nu}\Theta_{\nu\alpha})$,
where $d_\mu$ is a covariant derivative constructed from $\gamma_{\mu\nu}$.
Consequently, the restriction given by $S$-diffeomorphism invariance can
be presented in a $S$-diffeomorphism invariant way as
\begin{mathletters}%%
\begin{equation}
  d_\mu d_\nu \left(\epsilon^{\mu\alpha}\Theta_{\alpha\beta}\gamma^{\beta\nu}
  \right) = 0
\label{eq:33a}
\end{equation}
This implies that
\begin{equation}
  d_\mu \left(\gamma^{\mu\nu}\Theta_{\nu\alpha}\right) = \partial_\alpha
  \hbox{ (scalar)}
\label{eq:33b}
\end{equation}
\label{eq:33ab}%%
\end{mathletters}%%
which is the constraint on the divergence of the energy-momentum tensor
mentioned earlier.

Computing $\Theta_{\mu\nu}$ from (\ref{eq:25}) gives
\begin{equation}
  \Theta_{\mu\nu}  =  -\frac{1}{48\pi}
      \left(\partial_\mu \varphi \partial_v \varphi -
              \frac{1}{2} \gamma_{\mu\nu}
      \gamma^{\alpha\beta} \partial_\alpha \varphi \partial_\beta
                                   \varphi \right)
            - \frac{1}{24\pi}\left(d_\mu d_\nu \varphi - \frac{1}{2}
\gamma_{\mu\nu}
      \gamma^{\alpha\beta} d_\alpha d_\beta \varphi \right)
\label{eq:35}
\end{equation}
where $\varphi$ satisfies [compare (\ref{eq:6}) and (\ref{eq:6b})]
\begin{equation}
  \gamma^{\alpha\beta} d_\alpha d_\beta \varphi = \partial_\alpha
  \gamma^{\alpha\beta}\partial_\beta \varphi = {\cal R}
\label{eq:35b}
\end{equation}
Clearly $\Theta_{\mu\nu}$ is traceless, and one readily verifies that
\begin{equation}
  d_\mu \left(\gamma^{\mu\nu}\Theta_{\nu\alpha}\right) = \frac{1}{48\pi}
  \partial_\alpha {\cal R}
\label{eq:36}
\end{equation}
[compare (\ref{eq:7})], which is consistent with (\ref{eq:33ab}).

Finally we remark that even under $S$-diffeomorphisms ${\cal R}^\mu\/$
does not transform as a vector.  One finds from (\ref{eq:14}),
(\ref{eq:14c}) and (\ref{eq:24})
\begin{equation}
\delta_{\rm SD} {\cal R}^\mu
           =  L_f  {\cal R}^\mu + \epsilon^{\mu \nu} \partial_\nu \Delta_f
\label{eq:36a}
\end{equation}
where the vector field $f^\mu\/$ is now $-\epsilon^{\mu \nu}
\partial_\nu f\/$; thus here
$\Delta_f = 2 \partial_{+} \partial_{-} f - \tanh \frac{\beta}{2}
\left( e^\alpha \partial^2_{-} + e^{-\alpha} \partial^2_{+} \right) f$.

Although selecting between Weyl and $S$-diffeomorphism invariance on
the one hand or conventional diffeomorphism invariance on the other
remains a matter of arbitrary choice, as is seen from the fact that
the effective actions for the two options differ by local terms, the
following observations should be made in favor of the former.

\section{}
Up to now, the gravitational field $g_{\mu \nu}\/$ was a passive,
background variable.  Consider now the puzzles that arise when it is
dynamical; {\it i.e.}~$g_{\mu \nu}\/$ is varied.  With a single Bose
field, it is immediately established that the classical theory does
not possess solutions.  This is seen from the equation that follows
upon varying $g_{\mu\nu}$ in ${\cal L}$,
\begin{equation}
  \partial_\mu\phi\partial_\nu\phi-\frac{1}{2} g_{\mu\nu}
  g^{\alpha\beta}\partial_\alpha\phi\partial_\beta\phi = 0
\label{eq:37}
\end{equation}
which implies that
$g_{\mu\nu}\propto\partial_\mu\phi\partial_\nu\phi$, so that $g$
vanishes and $g^{\mu\nu}$ does not exist; alternatively $\phi$ must be
constant and $g_{\mu\nu}$ undetermined.  If there are $N$ scalar
fields, whereupon the effective action acquires the factor $N$, the
above difficulty is avoided, because $g_{\mu\nu} \propto
\sum^{N}_{i=1} \partial_\mu\phi^i\partial_\nu\phi^i$ need not be
singular.  Nevertheless (\ref{eq:37}) (with field bilinears replaced
by sums over the $N$ fields) requires the vanishing of positive
quantities $\sum^{N}_{i=1} \left\{\left(\gamma^{00}\dot{\phi}^i +
\gamma^{01}\phi'^i\right)^2 + \left(\phi'^i\right)^2 \right\}$ and
again only the trivial solution is allowed.

The quantum theory in Hamiltonian formulation also appears
problematic, in that the constraints of vanishing ${\cal E}$ and
${\cal P}$ cannot be imposed on states.  With one scalar field, the
momentum constraint requiring that $\phi'\Pi$ acting on states vanish
--- this is the spatial diffeomorphism constraint --- forces the state
functional in the Schr\"{o}dinger representation to have support only
for constant fields $\phi$.  (Equivalently one observes that a spatial
diffeomorphism invariant functional cannot be constructed from a
single, $x$-dependent field.)  With more than one field, this problem
is absent [a diffeomorphism invariant functional can involve $\int d x
\phi_1(x)\phi'_2(x)$] and the momentum constraint can be solved.
However, an obstruction remains to solving the energy constraint,
owing to the well-known Schwinger term (Virasoro anomaly) in the
$[{\cal E},{\cal P}]$ commutator, which gives a central extension that
interferes with closure of constraints: classical first-class
constraints become upon quantization second-class.
\begin{mathletters}%%
\begin{eqnarray}
 i \left[{\cal E}(x),  {\cal E}(y)\right]
           & = & i \left[{\cal P}(x),{\cal P} (y)\right]
                      =   \left({\cal P}(x)+{\cal P}(y))\delta'(x-y)\right)
                      \label{eq:38a} \\
 i \left[{\cal E}(x),  {\cal P}(y)\right]
           & = & \left({\cal E}(x) + {\cal E} (y)\right)\delta'(x-y)
                      - \frac{N}{12\pi}\delta^{'''}(x-y)
                      \label{eq:38b}
\end{eqnarray}
\label{eq:38ab}%%
\end{mathletters}%%
Note that all the above troubles, both in the classical theory and in
the Dirac-quantized Hamiltonian theory, revolve around diffeomorphism
invariance, not Weyl invariance.  Indeed the same troubles persists
for massive scalar fields, which are not Weyl invariant.

Thus when a quantum theory is constructed by a functional integral
(not by Hamiltonian/Dirac quantization) it is natural that it should
reflect problems with diffeomorphism invariance --- reducing it to
$S$-diffeomorphism invariance.  Weyl invariance on the other hand
could survive quantization.

\acknowledgements \hfil\break\noindent
I thank S. Deser, V. P. Nair, W. Taylor, and B. Zwiebach for comments.

\nonfrenchspacing
\section*{Added Note:~~~~}
Many of the results presented here on $S$-diffeomorphism invariance
have already appeared in D.~Karakhanyan, R.~Manvelyan and
R.~Mkrtchyan, {\it Phys.~Lett.}~{\bf B 329} (1994) 185.  Hence my
paper is not for publication.

\end{document}